\documentclass[runningheads]{llncs}
\usepackage[T1]{fontenc}
\usepackage{float}
\usepackage{cite}
\usepackage{array}
\usepackage{geometry} 
\usepackage{rotating}
\usepackage{ulem}
\usepackage{amsmath,amssymb,amsfonts}
\usepackage{algorithmic}
\usepackage{graphicx}
\usepackage{textcomp}
\usepackage{subcaption}
\usepackage{xcolor}
\usepackage{graphicx}
\usepackage{caption}

\makeatletter
\newcommand{\printfnsymbol}[1]{%
  \textsuperscript{\@fnsymbol{#1}}%
}
\newcommand{\printfnsymbolB}[1]{%
  \textsuperscript{%
    \ifcase#1 *\or †\or ‡\or §\or ‖\or ¶\else *\fi%
  }%
}
\makeatother
\begin{document}
\title{DiffNMR: Advancing Inpainting of Randomly Sampled Nuclear Magnetic Resonance Signals}
%
%
\author{Sen Yan\thanks{equal contribution}\and
Fabrizio Gabellieri\printfnsymbol{1} \and
Etienne Goffinet\printfnsymbol{1} \and
Filippo Castiglione \and
Thomas Launey
}
\authorrunning{S. Yan et al.}
%
\institute{Biotechnology Research Center, Technology Innovation Institute, P.O. Box 9639, Abu Dhabi, United Arab Emirates,
\email{\{Sen.Yan, Fabrizio.Gabellieri, Etienne.Goffinet, Filippo.Castiglione, Thomas.Launey\}@tii.ae}
}
\maketitle              
\begin{abstract}
Nuclear Magnetic Resonance (NMR) spectroscopy leverages nuclear magnetization to probe molecules' chemical environment, structure, and dynamics, with applications spanning from pharmaceuticals to the petroleum industry.\quad Despite its utility, the high cost of NMR instrumentation, operation and the lengthy duration of experiments necessitate the development of computational techniques to optimize acquisition times. Non-Uniform sampling (NUS) is widely employed as a sub-sampling method to address these challenges, but it often introduces artifacts and degrades spectral quality, offsetting the benefits of reduced acquisition times.
In this work, we propose the use of deep learning techniques to enhance the reconstruction quality of NUS spectra. Specifically, we explore the application of diffusion models, a relatively untapped approach in this domain.\quad Our methodology involves applying diffusion models to both time-time and time-frequency NUS data, yielding satisfactory reconstructions of challenging spectra from the benchmark Artina dataset. This approach demonstrates the potential of diffusion models to improve the efficiency and accuracy of NMR spectroscopy as well as the superiority of using a time-frequency domain data over the time-time one, opening new landscapes for future studies.
\end{abstract}
\section{Introduction}

Nuclear Magnetic Resonance (NMR) spectroscopy is a cornerstone in modern analytical chemistry, offering unparalleled insights into molecular structure, dynamics, and interactions.
During an NMR analysis, a sample containing atoms with specifically sensitive nuclei (possessing a spin
property) is placed into a strong magnetic field. When subjected to radio frequencies, the nuclei absorb and re-emit energy at specific frequencies, a phenomenon known as resonance. After recording the resonance across a frequency range, experts or specialized models then analyze this spectrum to help decipher the target molecule structure.


\begin{figure}[tb]
    \centering
    \includegraphics[width=0.6\linewidth]{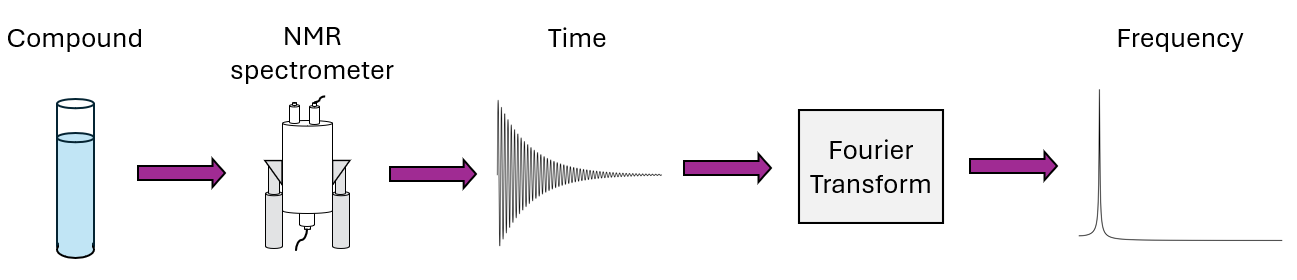}
    \caption{A 1D NMR experiment records the nuclei resonance in response to a radiofrequency pulse.}
    \label{fig:1d_exp}
\end{figure}
NMR spectroscopy can be conducted in one dimension (1D), two dimensions (2D), or more dimensions. Each provides different levels of information. 1D NMR is the simplest form, where a resonance time series (time domain) called Free Induction Decay (FID) is obtained by applying a single radiofrequency pulse (\textit{c.f.}, Fig.~\ref{fig:1d_exp}). After Fourier transform, The resonance is turned into a spectrum (frequency domain). This method provides fundamental data about the chemical environment of specific nuclei, such as the number of protons and their electronic surroundings. However, when molecular complexity increases, 1D NMR may not be sufficient to resolve overlapping signals.
\begin{figure}[tb]
    \centering
    \includegraphics[width=0.6\linewidth]{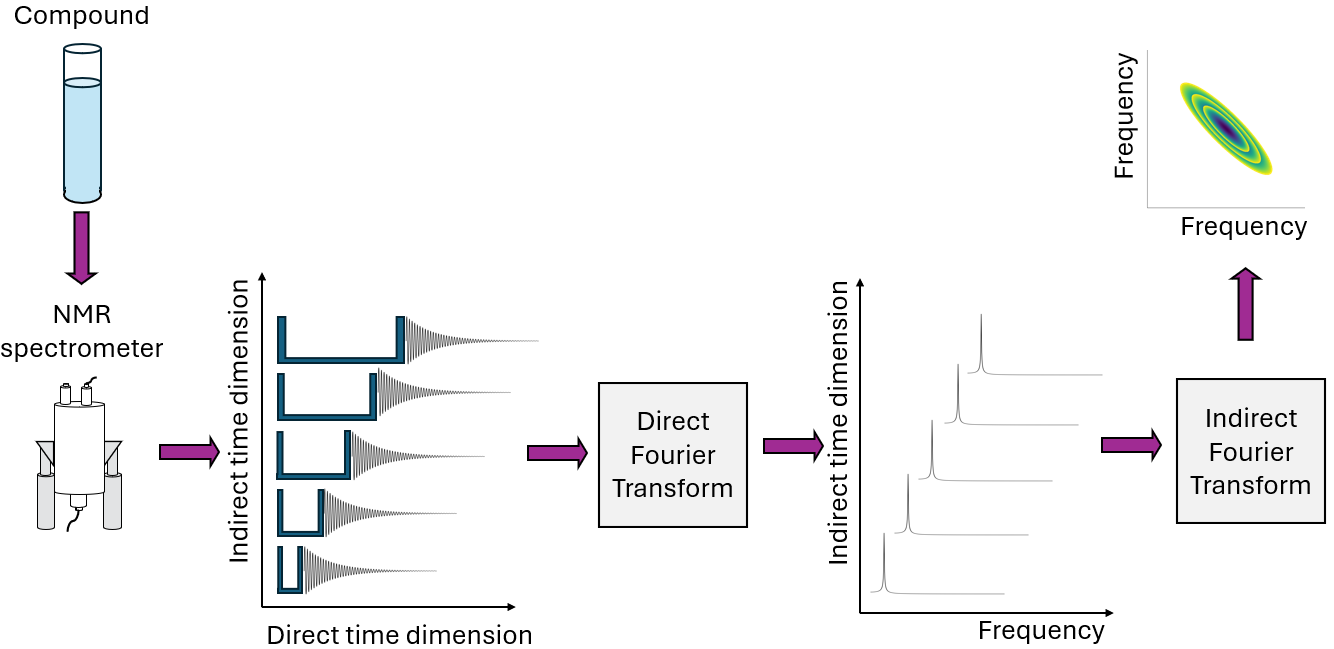} 
    \caption{A 2D NMR spectrum is obtained by applying a series of radiofrequency pulses and evolution time delays to a sample (here, five different values), and apply a 2D Fourier transform to obtain a 2D frequency spectrum.}
    \label{fig:2d_fully_sampled}
\end{figure}

Higher-dimensional NMR spectroscopy can help alleviate this issue and show interactions between nuclei (\textit{e.g.}, through-bond or through-space couplings). In 2D NMR, multiple 1D experiments are conducted sequentially, with variations in the timing and sequence of pulses (illustrated in Fig.~\ref{fig:2d_fully_sampled}), which generates a 2D resonance map (time-time domain). 

In this record, the x-axis (called direct dimension) corresponds to the frequency detected directly after applying the final pulse sequence, similar to 1D NMR. The indirect dimension is obtained by systematically varying the time between pulses in a series of experiments, known as the evolution time. The final spectrum (frequency-frequency domain) is then obtained with 2D Fourier transform. 2D NMR offers a more detailed view of molecular structures, enabling the identification of connectivity between atoms within a molecule, which is especially useful for analyzing large molecules. 

\begin{figure}[tb]
    \centering
    \includegraphics[width=0.6\linewidth]{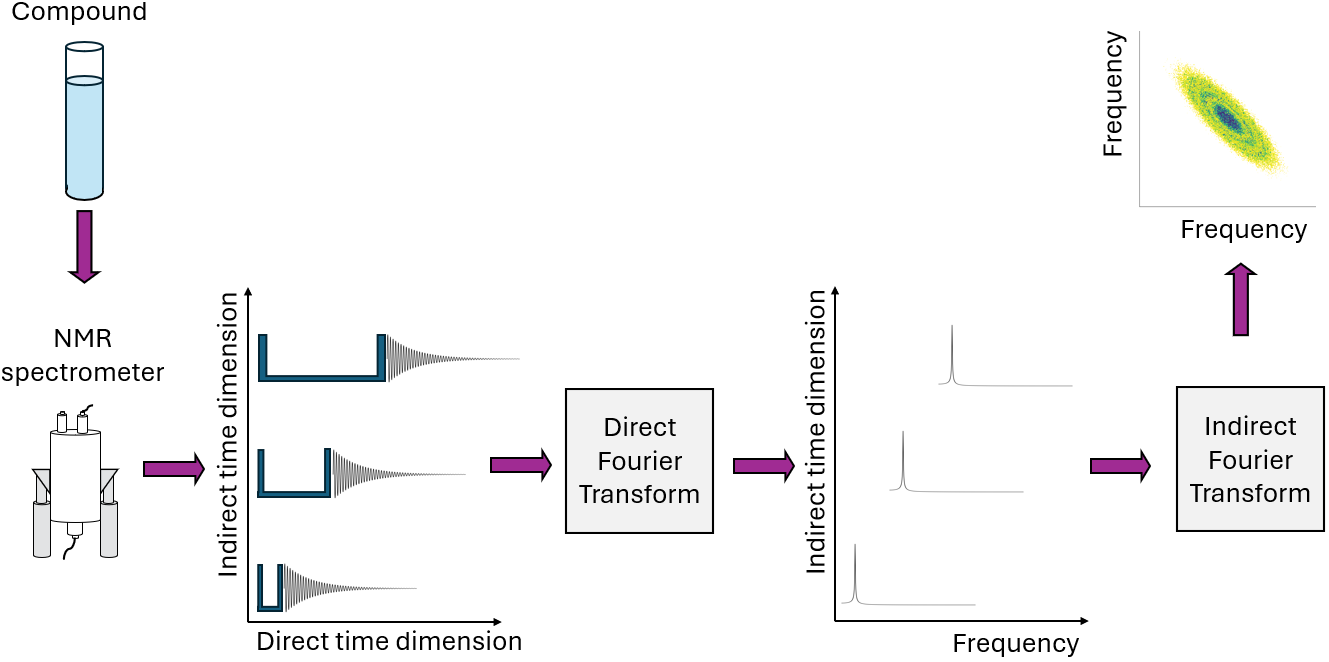}
    \caption{Non-Uniform Sampling: some of the evolution times (here: the second and fourth) are skipped.}
    \label{fig:2d_undersampled}
\end{figure}

2D NMR higher spectral resolution comes at the expense of acquisition speed: 2D experiments take 10 to 1000 times more time than a 1D one \cite{delaglio2017non}. Consequently, methods are paramount to cut acquisition times and preserve experiments' resolution.
NUS techniques \cite{barna1987exponential,schmieder1994improved} reduces acquisition time by randomly skipping some evolution time values, as illustrated in Fig.\ref{fig:2d_undersampled}, usually following a Poisson-gap sampling \cite{hyberts2010poisson} or random sampling \cite{delaglio2017non}. In our work, we focused on the latter method.

One drawback of using NUS techniques is the production of artifacts in the collected spectra \cite{zhan2024fast}. Consequently, pursuing approaches for ameliorating this problem has been an active field of research.

By postulating the undersampling of the FIDs, potentially infinite ways of reconstructing the original data are available.
Compressed Sensing (CS) imposes the constraint of minimal $l_{0}$ norm (number of nonzero elements) over the frequency domain data to select a solution. In practice, it minimizes a convex penalty function using the $l_{1}$ norm of the spectrum \cite{kazimierczuk2011accelerated}. Despite its strong performance in reducing acquisition times, CS can introduce artifacts if sparsity assumptions are incorrect and often requires computationally intensive reconstruction.

Low-rank approximation (LR) \cite{qu2015accelerated} is an alternative reconstruction method that reconstructs the missing data by approximating the matrix with the lowest possible rank that still captures the essential features of the observed data. However, the literature has shown that the LR method can produce  sensitive to noise and incorrect rank assumptions \cite{zhan2024fast},\cite{qu2015accelerated}.

In recent years, several Deep Learning (DL) alternatives have been proposed, showcasing various methodologies and their applications in improving NMR data analysis and interpretation.
Qu et al. \cite{qu2020accelerated} introduced a dense Convolutional Neural Network (CNN) \cite{lecun2015deep} with a spectrum consistency block for Poisson-Gap NUS data, achieving promising results for 2D and 3D spectra. Zhan et al. \cite{zhan2024fast} employed SEPSNet, an attention-based network, to enhance pure shift NMR reconstruction from 2D resonance map. Zheng et al. \cite{zheng2022fast} developed PS-ResNet, a ResNet-based model, for denoising 1D spectra. Karunanithy et al. \cite{karunanithy2021fid} applied dilated convolutions through Fidnet, inspired by WaveNet, for synthetic NMR data analysis.

Despite the advancements, these deep learning methods were primarily trained on simulated datasets and have not consistently surpassed the performance of the established CS baseline for NUS reconstruction. This limitation underscores the importance of moving beyond synthetic data to real-world datasets with greater variability and complexity.

The recent introduction of diffusion models for NMR spectral reconstruction provides a promising alternative. Unlike traditional deep learning techniques, diffusion models are specifically trained to remove noise from data and reconstruct missing parts. By iteratively refining the reconstruction based on the entire data distribution, diffusion models are well-suited for \emph{real} NMR data, where noise and sampling irregularities are more prominent. This approach justifies their application to real datasets, offering the potential for improved robustness and accuracy in challenging NUS scenarios.

\begin{figure}[tb]
    \centering
    \includegraphics[width=0.8\linewidth]{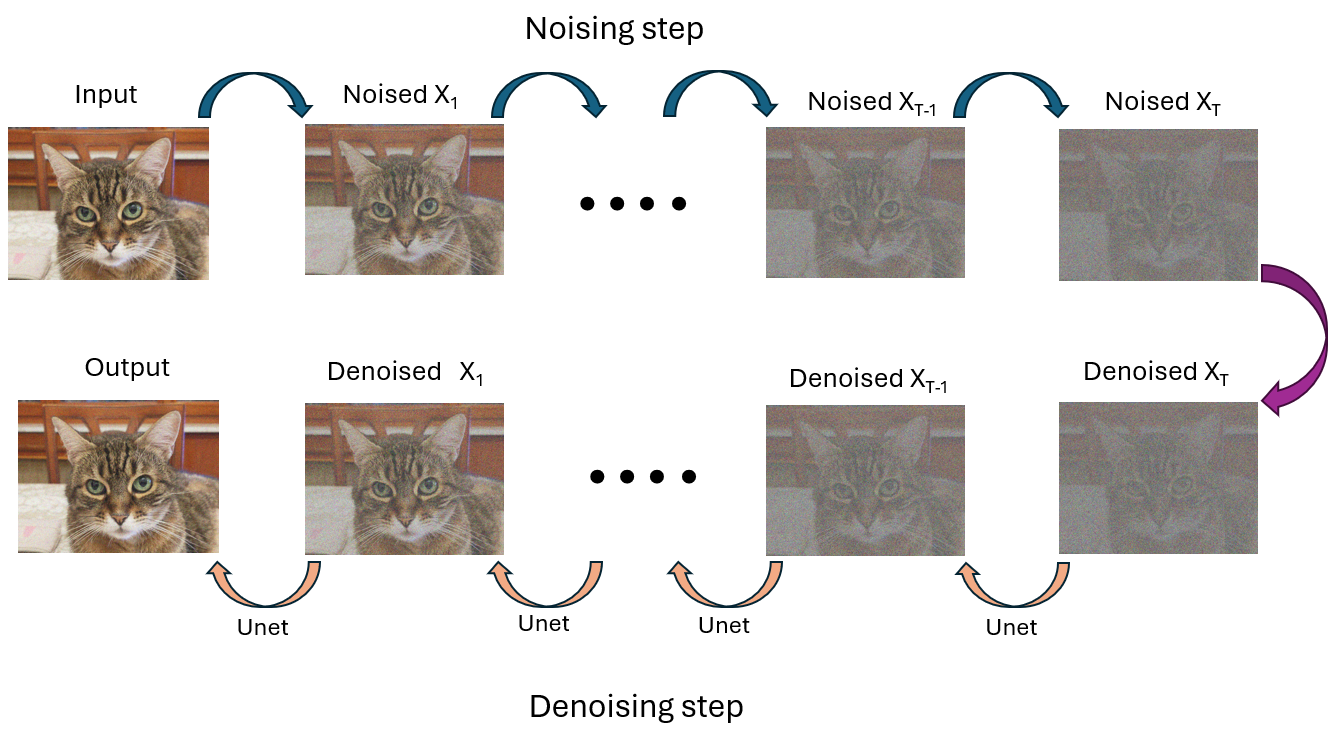}
    \caption{Diffusion steps: in the noising step (forward pass), the input is progressively corrupted. The Unet model is trained to predict the added noise (backward pass).}
    \label{fig:diffusion_cat}
\end{figure}

In this work, we consider the protein NMR dataset from \cite{klukowski2022rapid} rather than simulated datasets to ensure that our results are grounded in realistic scenarios that exhibit specific noise. 

Accordingly, we provide a comprehensive account of all transformations applied to the data, ensuring that others in the field can accurately reproduce our implementation. 
Specifically, we extract time-time and time-frequency domain representations from the dataset by applying an inverse Fourier transform, either on both dimensions or just on the indirect one. This approach allows us to explore the use of diffusion models in spectra reconstruction across two distinct datasets, each corresponding to a specific experimental setting.

In particular, for each domain (time-time and time-freq), we train a diffusion model on a denoising task and a conditioned diffusion model on an inpainting task. During inference, we evaluate the models' ability to reconstruct the spectra under varying random NUS masking levels in the appropriate domain data.

\section{Materials and methods}

\subsection{Data description}
In recent literature, diffusion models for image generation have been trained on hundreds of millions of images \cite{schuhmann2021laion,rombach2022high}. In comparison, 2D NMR protein spectra dataset is a rare resource, with few public repositories available. While the standard solution in the literature \cite{zhan2024fast},\cite{qu2020accelerated},\cite{karunanithy2021fid}, \cite{zheng2022fast} is to train on synthetic spectra, we chose in this study to consider only real-life data, which contains more realistic noise and signal.

The training dataset is based on the 100-protein NMR spectra dataset \cite{klukowski2022rapid},\cite{klukowski2024100}, which consists of 1329 spectra (2D/3D/4D) describing 100 proteins and sampled from 600 to 950 MHz NMR machines. 

In addition to extracting the original 2D NMR spectra from this dataset, additional 2D spectra are generated by flattening the higher-dimensional spectra on 2D representations. This process is not unfamiliar to NMR users, as it is commonly performed to visualize higher-dimension spectrum. In total, this generates more than 3500 samples.

\begin{figure}[tb]
    \centering
    \includegraphics[width=0.8\linewidth]{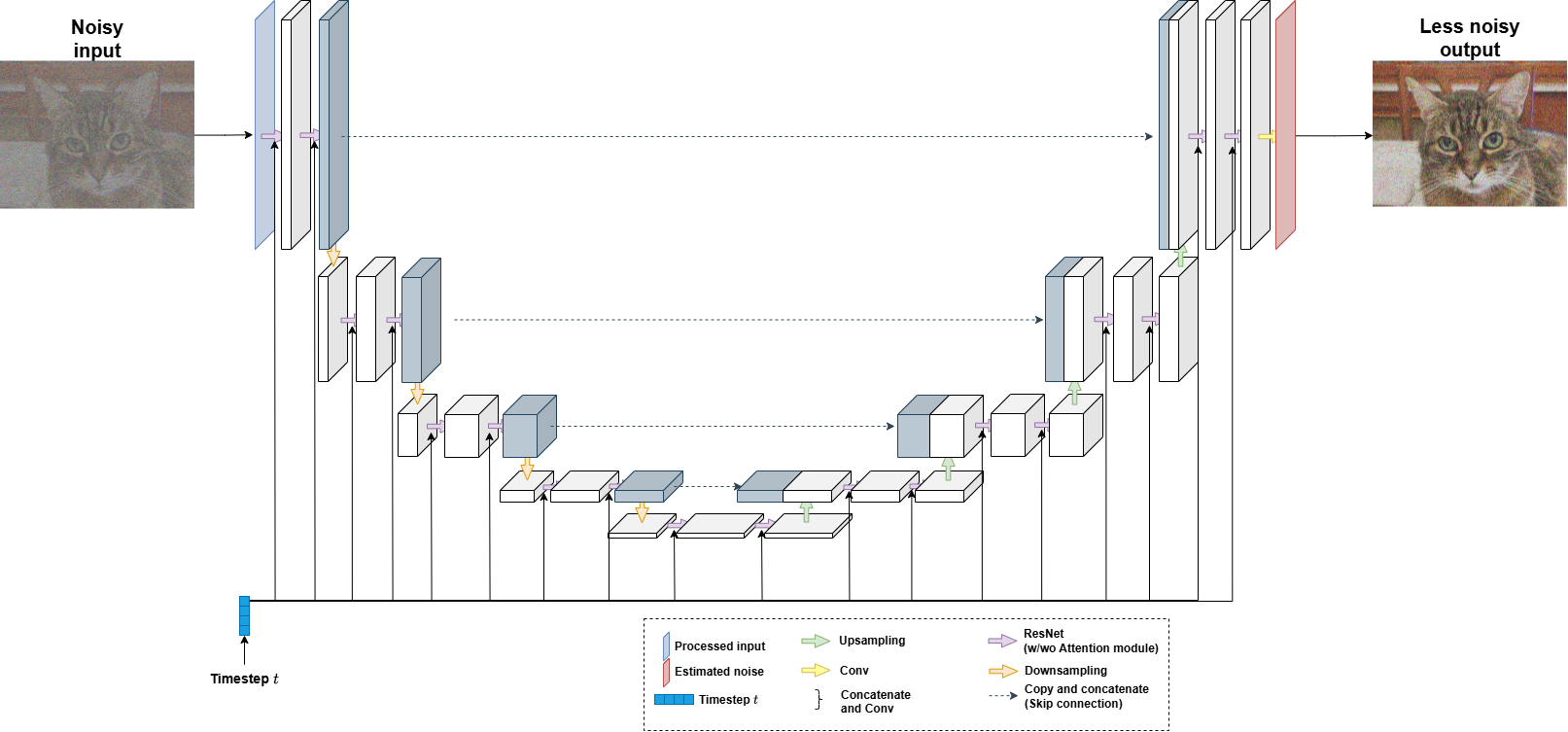}
    \caption{UNet model for noise prediction. }
    \label{fig:unet}
\end{figure}

\subsection{Brief introduction to diffusion models}
The explosion of interest in generative models came in the 2010s, as deep learning techniques became matured. 
Generative models learn the underlying patterns and structures of the data, enabling them to produce novel instances. These models can generate a wide range of data types, such as images \cite{ganimation,yan2023combining,imga,yan2023personalizing}, text \cite{bert,gpt,touvron2023llama}, and audio \cite{van2016wavenet,diffusion-audio}, and are capable of tasks like data augmentation \cite{sr3,srdiff}, content creation \cite{touvron2023llama}, and style transfer \cite{karras2019style}.
This period saw the development of key generative models such as Generative Adversarial Networks (GANs) \cite{gan}, Variational Autoencoders (VAEs) \cite{vae}, and more recently Diffusion models \cite{ho2020denoising}.

Denoising Diffusion Probabilistic Models (DDPM) \cite{ho2020denoising}, are a class of neural networks originally inspired by nonequilibrium thermodynamics processes \cite{sohl2015deep}. These models are trained by progressively destroying data with noise injection and then learning to reverse the process for sample generation.

In the "noising" step (called forward pass), the input is iteratively corrupted by adding scheduled Gaussian noise. Each data point will approximate a standard normal variable after a sufficiently high number of noising iterations. In the second ("denoising" or backward pass) phase, a UNet \cite{ronneberger2015u} model learns to predict the noise added to the data points at each timestep, allowing for the reconstruction of the original data distribution. The two steps are displayed in Fig. \ref{fig:diffusion_cat}. A representation of the Unet architecture is shown in Fig.\ref{fig:unet}.

Data can then be generated by sampling from a standard normal distribution and iteratively denoising it using the trained UNet.
\begin{figure}[tb]
    \centering
    \includegraphics[width=0.75\linewidth]{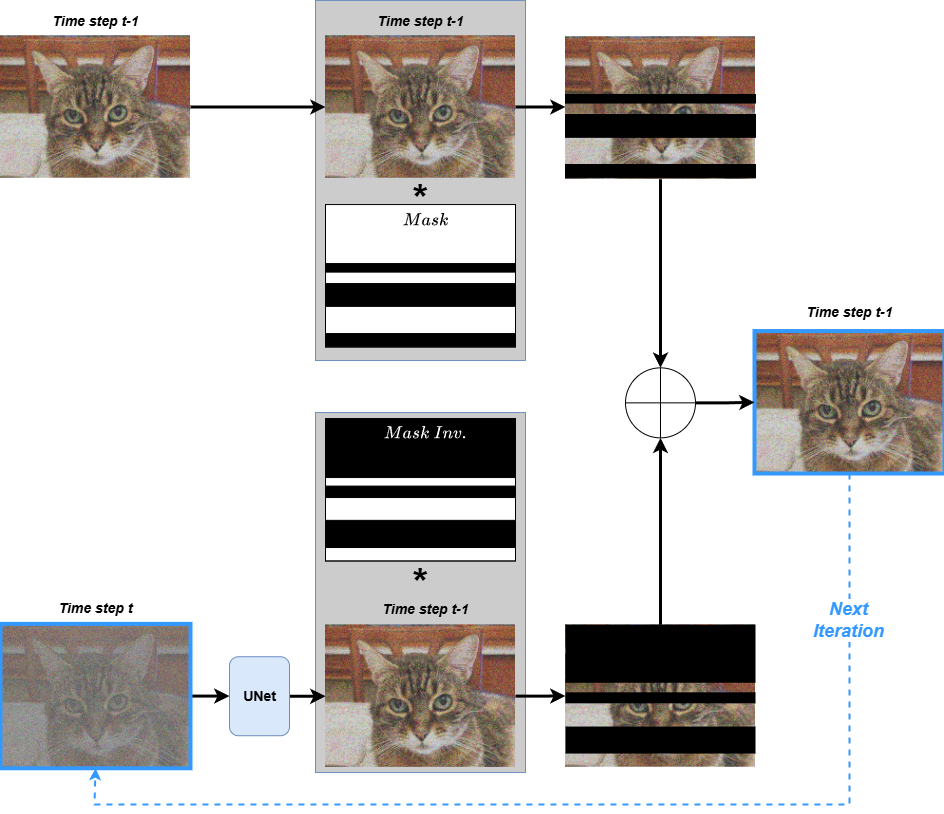}
    \caption{Inpainting using a denoising model with repaint schedule}
    \label{fig:denoise_repaint}
\end{figure}

\begin{figure}[tb]
    \centering
    \includegraphics[width=0.8\linewidth]{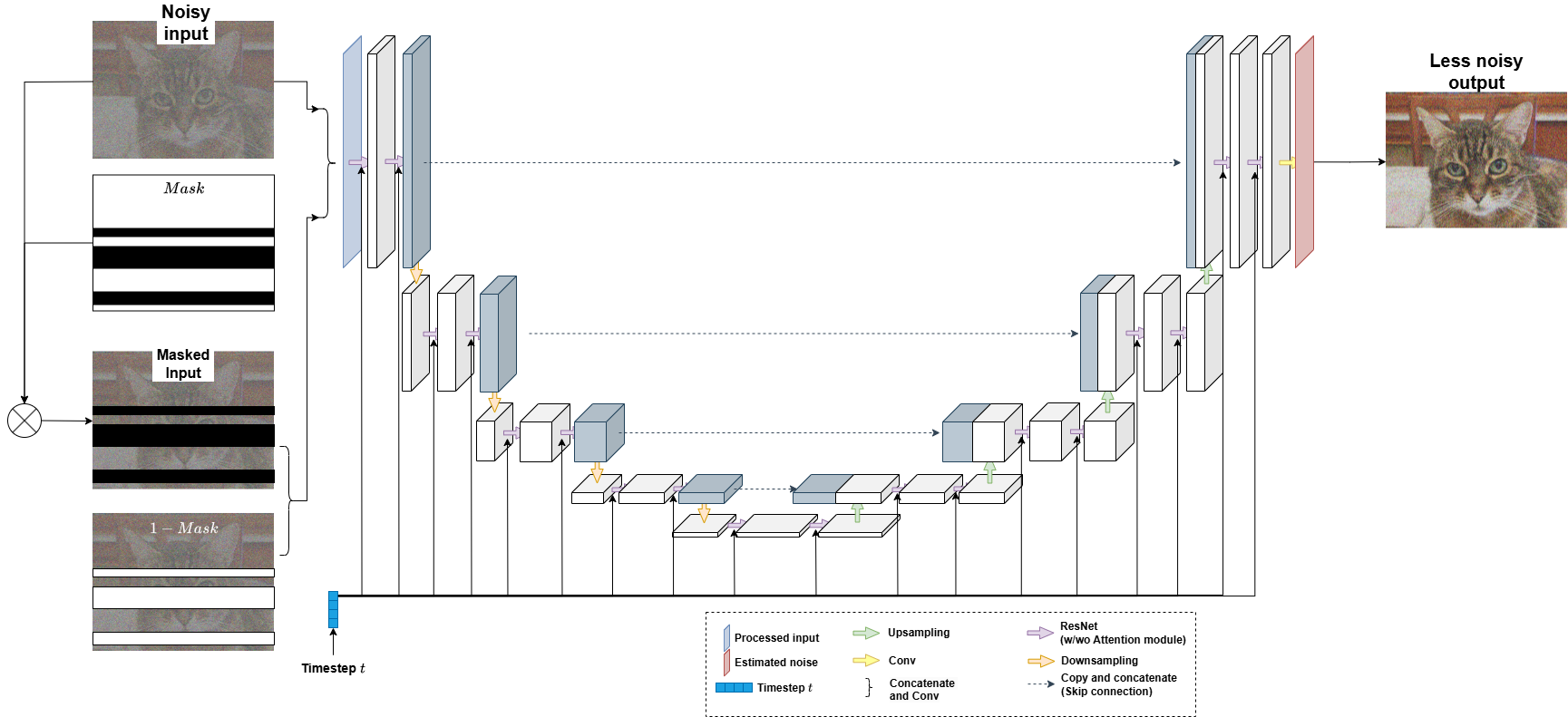}
    \caption{A conditioned UNet for inpainting.}
    \label{fig:inpainting}
\end{figure}

DDPM can be adapted to the inpainting task, allowing for the reconstruction of missing parts of an image.
A masked image is provided, and noise is added to the known and unknown parts. The model can then reconstruct the missing parts during the denoising phase by using the context provided by the unmasked regions: those guide the reconstruction and allow for contextually accurate predictions of the missing areas.

Inpainting with diffusion models can be performed in several ways. The first approach is to train a denoising model and use a repainting pipeline \cite{lugmayr2022repaint}, where, at every denoising step, the unmasked part of the spectrum is re-used and super-imposed on the denoised output. This approach is shown in Fig \ref{fig:denoise_repaint}.

Another version consists of training a conditioned diffusion model (\textit{c.f.}, Fig. \ref{fig:inpainting}). Differing from the denoising model, the input of the conditioned diffusion model is the concatenation of the mask (as an additional channel) and the original sample (see Fig.~\ref{fig:inpainting}).

\subsection{Diffusion models for spectra reconstruction}
Our study aims to explore the effectiveness of diffusion models in reconstructing 2D spectra from random NUS data. From this perspective, the problem can be seen as an inpainting task, considering that the data samples are 'masked' in the time-time domain and require completion. Compared to standard inpainting on images, this application's specificity is that rows mask 2D inputs since NUS involves skipping entire 1D observations.

\begin{figure}[tb]
    \centering
    \includegraphics[width=0.8\linewidth]{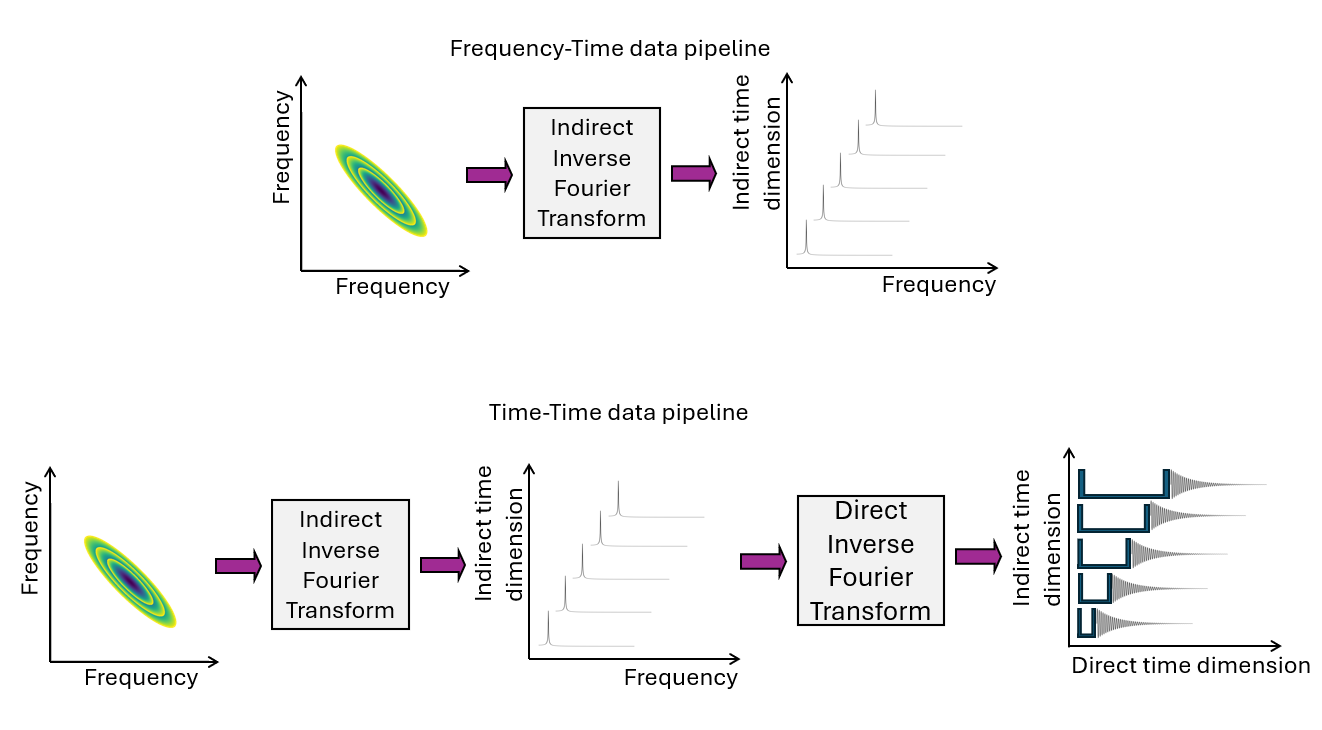}
    \caption{Above, the generation of the time-frequency domain data. Inverse Fourier transform is applied to the indirect dimension.
    Below, the time-time domain: Inverse Fourier transform is applied to the indirect dimension and then to the direct one.}
    \label{fig:tt_vs_ft}
\end{figure}

In our comprehensive study, we compare the denoising-based (Fig.~\ref{fig:denoise_repaint}) and conditioned diffusion-based (Fig.~\ref{fig:inpainting}) approaches. Furthermore, we conduct reconstruction experiments in both the time-time domain and the time-frequency domain (Fig.~\ref{fig:tt_vs_ft}), which helps assess the impact of the representation domain on the reconstruction quality. After reconstruction, we apply the appropriate Fourier transforms to obtain the frequency-frequency data (\textit{i.e.}, the spectrum). The training and evaluation datasets are obtained from the Artina dataset \cite{klukowski2022rapid,klukowski2024100} by applying inverse Fourier transform on the appropriate dimensions, depending on the case. In total, we trained four different models: two denoising-based models (Fig.~\ref{fig:denoise_repaint}) which we denote as $D-TT$ (Denoising - Time-Time domain) and $D-TF$ (Denoising - Time-Frequency domain), and two conditioned models (Fig.~\ref{fig:inpainting}) which we denote as $I-TT$ (Inpainting - Time-Time domain) and $I-TF$ (Inpainting - Time-Frequency domain). 
\begin{figure}[tb]
    \centering
    \begin{subfigure}[b]{0.45\linewidth}
        \centering
        \includegraphics[width=\linewidth]{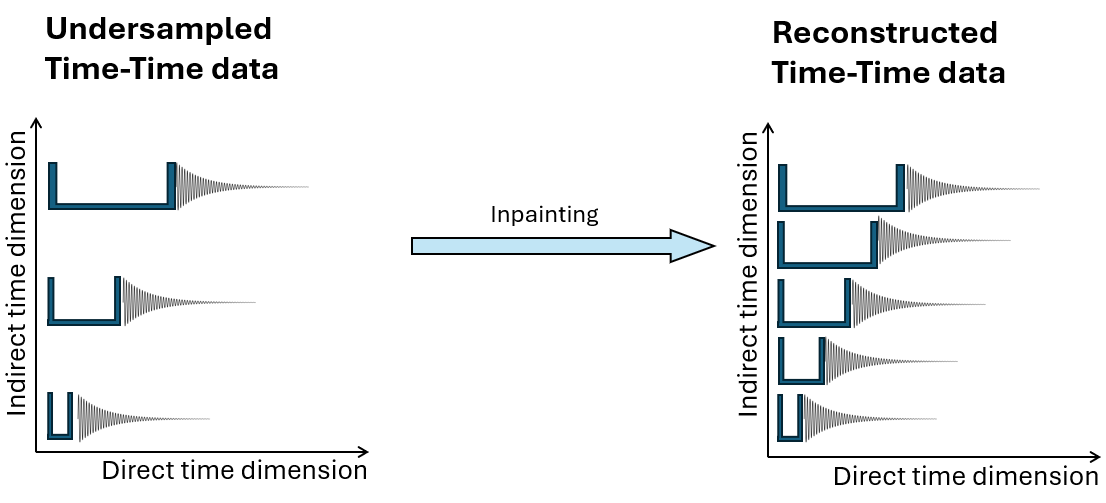}
        \caption{The experiment in time-time domain.}
        \label{fig:time_time}
    \end{subfigure}
    \hfill
        \begin{subfigure}[b]{0.45\linewidth}
        \centering
        \includegraphics[width=\linewidth]{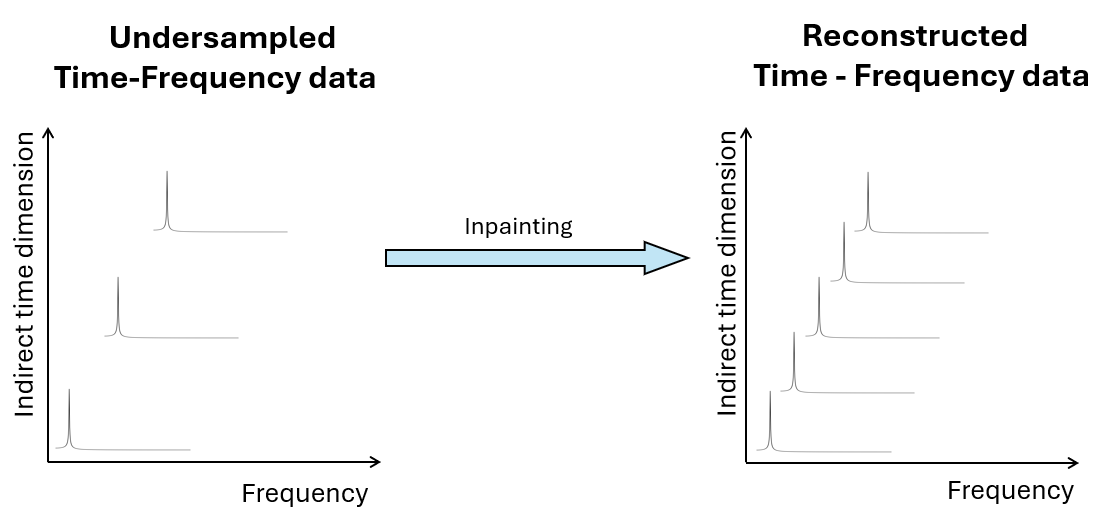}
        \caption{The experiment in time-frequency domain.}
        \label{fig:freq_time}
        \end{subfigure}
    \caption{We set two experiments shown in \ref{fig:time_time}: time-time domain and in \ref{fig:freq_time} time-frequency domain. In each experiment, we trained a denoising model and an inpainting model to reconstruct the data.}
    \label{fig:tasks}
\end{figure}

\section{Experiments}
\subsection{Training}
We carry out two sets of experiments, shown in Fig. ~\ref{fig:time_time} and Fig.~\ref{fig:freq_time}, employing our four models, \textit{i.e.}, the diffusion models for the time-time domain and time-frequency domain as well as the conditioned diffusion models for the time-time domain and time-frequency domain.

The preprocessing of the data involves the original Artina data resizing, followed by the appropriate application of inverse Fourier transform (either on the indirect and direct dimension or just on the indirect one). The result is a complex number tensor stored with two channels.

We train the models with Mean Squared Error (MSE) loss on the preprocessed Artina dataset \cite{klukowski2022rapid,klukowski2024100} with a train-validation split ratio of (0.88, 0.12) and always select the model that minimizes the validation loss to avoid overfitting. Both the denoising and inpainting models use a DDPM scheduler for the noising step. In the case of inpainting training, we add a mask that randomly masks entire rows of the data (which, depending on the experiment setting, are either made of time or frequency points).

\subsection{Metrics and Baselines}

The reconstructed spectrum quality is evaluated with global and local effectiveness metrics.

\textbf{Global metrics.}
We measure the NUS global reconstruction error with the Mean Square Error (MSE), the coefficient of determination ($R^2$), and the ratio of the signal-to-noise ratio (SNR) between the reconstruction and the original spectrum. MSE quantifies the average squared difference between the reconstructed spectrum and the true spectrum, serving as a direct measure of reconstruction accuracy. $R^2$, on the other hand, evaluates the proportion of variance in the true spectrum explained by the reconstructed spectrum. The ratio of SNR between the reconstruction and the original spectrum provides a quantitative measure of how well the reconstructed spectrum preserves the signal relative to the original spectrum. A ratio close to 1 indicates that the reconstruction has maintained the original signal-to-noise ratio. By comparing the SNR, we can evaluate the effectiveness of the reconstruction method in suppressing noise without distorting the original signal. 

\textbf{Local metrics.}
Because an NMR spectrum informative value lies mainly in the position and amplitudes of the peaks, we also use local metrics focused on these peaks. Specifically, we use nmrglue to locate and select the peaks of the original spectrum and the reconstructed spectrum and then apply the Linear Sum Alignment algorithm \cite{crouse2016implementing} to match them. From this, we investigate the hallucination ratio of the peaks, which refers to the proportion of peaks in a reconstructed spectrum that do not correspond to any actual peaks in the reference spectrum. These peaks (artifacts) are the false detection peaks introduced during inpainting.
 
Our experiments cover a range of masking proportions, from 0.5 to 0.95, to examine the performance evolution with increasing NUS mask data. The displayed score is the average over 5 runs.

In the absence of the code and model weights of the deep learning alternatives \cite{qu2020accelerated}, we compare our diffusion model against the CS \cite{kazimierczuk2011accelerated} and LR \cite{qu2015accelerated} baselines.

\textbf{Compressed Sensing} (CS) \cite{kazimierczuk2011accelerated} exploits the sparsity of signals in a transformed domain to reconstruct high-resolution spectra from undersampled data, reducing acquisition time. The reconstruction problem is typically formulated as an $\ell_1$-norm minimization and solved using algorithms like Iterative Soft Thresholding (IST) or Alternating Direction Method of Multipliers (ADMM) to ensure accurate signal recovery while preserving sparsity.

\textbf{Low-Rank approximation} (LR) \cite{qu2015accelerated} leverages the observation that NMR spectra often exhibit a low-rank structure when represented as matrices. In practice, it consists of a matrix reduction method (Singular Value Decomposition) applied to a Hankel matrix representation of the spectrum.

\subsection{Evaluation set}
As an evaluation set, we apply our method to 10 compounds from the Artina dataset, which were not included in the training or validation splits. These compounds have Protein Data Bank (PDB) \cite{berman2003announcing} codes 2KBN, 2KCD, 2KHD, 2KJR (CBCANH), 2KJR (N15HSQC), 2KZV, 2L9R, 2LFI, 2LTM, and 2MK2.


\begin{figure}[tb]
     \centering
     \begin{subfigure}[b]{1\textwidth}
         \centering
         \includegraphics[width=\textwidth]{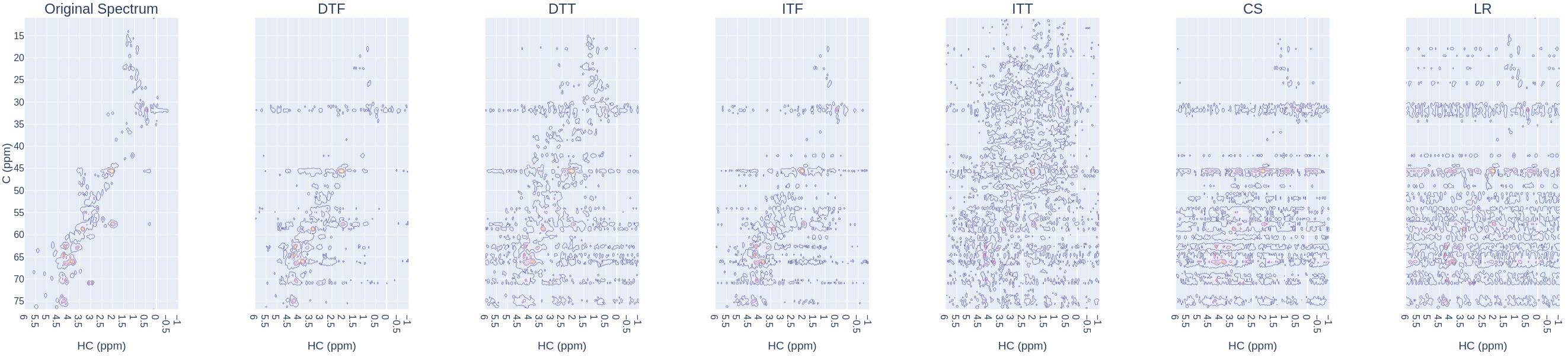}
     \end{subfigure}
     \hfill
     \begin{subfigure}[tb]{1\textwidth}
         \centering
         \includegraphics[width=\textwidth]{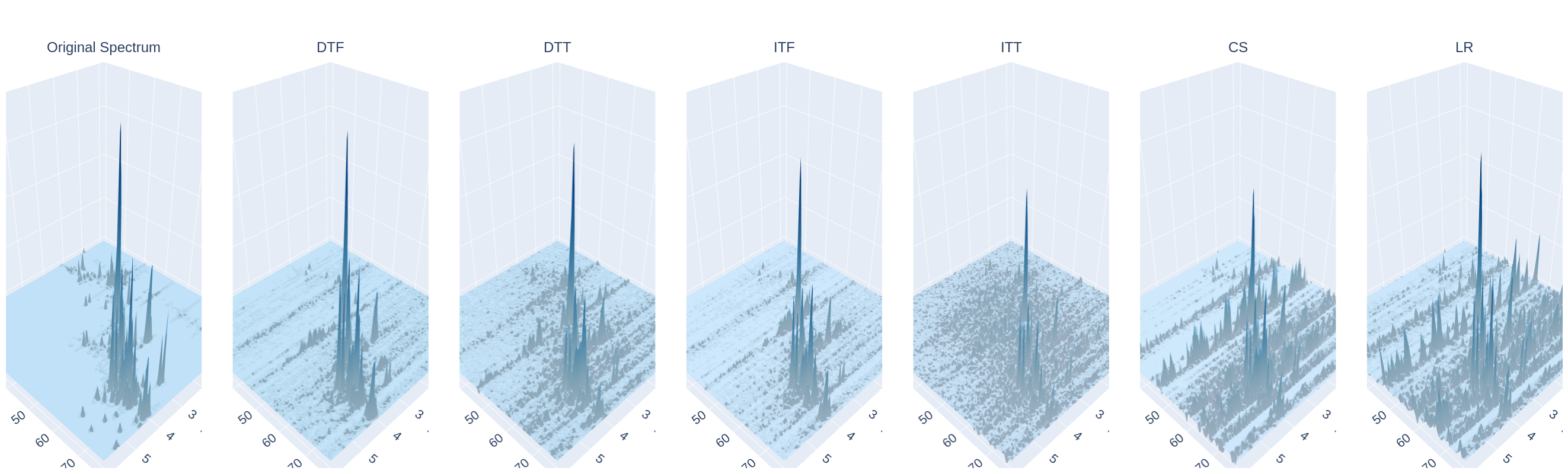}
     \end{subfigure}
     \caption{Reconstruction of spectrum: 2LFI - C13NOESY@aliphatic - Nuclei C HC at 30\% NUS (70\% masked). We illustrate the contours (top) and the 3D surface (bottom) of the spectra: (from left to right) Original spectrum, Reconstruction by DiffNMR ($D-TF$), Reconstructions by $D-TT$, $I-TF$, $I-TT$, as well as the baselines Compressed Sensing (CS) and Low-Rank Approximation (LR) for comparison. Please zoom in for better observations.}
     \label{fig:reconstruction_2K5V}
\end{figure}

Figures~\ref{fig:reconstruction_2K5V} presents a detailed comparison of the performance metrics across the different methods on our test set. The proposed method, in all its versions, consistently outperformed both CS and LR across all metrics.



\begin{figure}[tbp]
    \centering
    \includegraphics[width=0.9\linewidth]{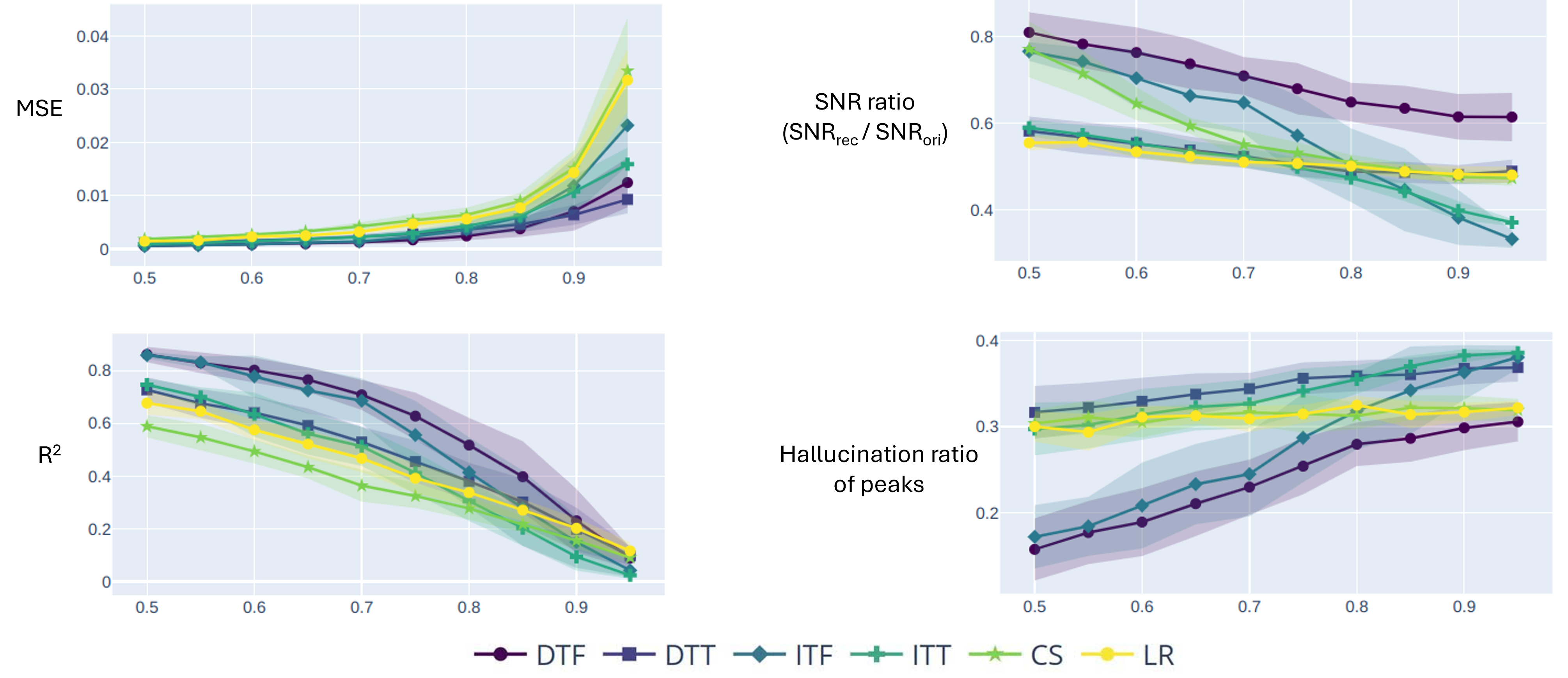}
    \caption{Global metrics for different masking ratios: overall the mean square error, the coefficient of determination, and the ratio of SNR between the reconstruction and the original spectrum. Local metrics for different masking ratios: the hallucination ratio of peaks. Note that the X-axis represents the percentage of masking.}
    \label{fig:global_scores}
\end{figure}

\subsection{Results}

\textbf{MSE}:
Figure~\ref{fig:global_scores} illustrates the global MSE for each method. All models maintain relatively stable and low error from 50\% to 80\% masking. MSE begins to increase significantly after 80\% of masking (i.e., 20\% NUS). CS and LR show the most pronounced error growth in the higher-masking ratio. DTF and DTT perform optimally, with minimal error growth even in a higher-masking ratio.
This reduction in error highlights the effectiveness of the proposed approach in accurately reconstructing the overall spectrum, which can also be observed in Figure~\ref{fig:reconstruction_2K5V}.

\textbf{Coefficient of determination, $R^2$}:
$R^2$ decreases as the percentage of masking increases, generally showing a decline in the model's ability to explain the variance in the data.
$D-TF$ and $I-TF$ outperform the others from 50\% to 75\% of masking. However, the SNR ratio of $I-TF$ drops quickly after 70\% of masking.
Figure~\ref{fig:global_scores} confirms the superiority of model $D-TF$. This suggests that overall and for every masking proportion, this model is more successful in capturing the variance of the reference spectra. The other model versions can outperform CS and LR from 50\% to 75\% masking.

\textbf{$SNR_{rec}/SNR_{ori}$}:
As shown in Figure \ref{fig:global_scores}, the ratio decreases as the percentage of masking increases, indicating a reduction in signal quality with more extensive masking.
$D-TF$ outperforms all the other models as well as the state-of-the-art methods. That is to say, compared with the others, the reconstruction by $D-TF$ preserves more information and well suppresses noise (see Figure~\ref{fig:reconstruction_2K5V}).

\textbf{Hallucination ratio of peaks}:
In terms of the reconstructed peaks, the hallucination ratio of peaks shown in Figure\ref{fig:global_scores} illustrates the superiority of $D-TF$ compared to the others. Indeed, before 80\% of masking, $D-TF$ and $I-TF$ generate fewer false-detection peaks than the other proposed models as well as the SOTA methods (i.e., CS and LR). $D-TF$ achieves the lowest hallucination ratio, even under high masking percentages.

Overall, the experimental results highlight that the $D-TF$ model consistently outperforms all other approaches across every metric, establishing it as the most robust and reliable method. Notably, when the masking ratio is below 75\%, $I-TF$ demonstrates comparable performance to $D-TF$ across all metrics. This finding underscores the advantage of processing data in the time-frequency domain over the time-time domain for the NUS inpainting task.

\section{Conclusion}
In this work, we present an alternative approach to tackle NUS data reconstruction. 
Our research leveraged diffusion models to reconstruct time-time and time-frequency domain undersampled data from the benchmark Artina protein dataset \cite{klukowski2022rapid,klukowski2024100}. We conducted experiments in both time-time and time-frequency domains and compared a diffusion model on a denoising task and a conditioned diffusion model on an inpainting task.
Using a well-cured nonsynthetic benchmark dataset ensured reproducibility and adherence to real-life settings.
We showed the robustness of our approach on an independent test dataset by showcasing several metrics that capture our model's global (MSE, $R^2$, SNR ratio) and local (Hallucination ratio of peaks) ability to reconstruct undersampled spectra. 
The results obtained on time-frequency domain data prove the validity of our novel approach and pave the way for further research using this data domain.
Among the possible future research directions, we envision using diffusion models in the context of Poisson-Gap NUS data.

\begin{credits}
\subsubsection{\ackname} We would like to express our sincere gratitude to our colleagues Ryan Young, Idir Malki, Lydia Gkour and Laurence Jennings for their valuable insights and support throughout this project. Their feedback and discussions were instrumental in refining our approach. 
\end{credits}
%
%
%
%
\bibliographystyle{splncs04}
\bibliography{reference_paper}




\end{document}